\documentclass[prb,floatfix,twocolumn,showpacs,superscriptaddress]{revtex4}
\usepackage{graphicx}
\usepackage{amsmath}
\def\bea{\begin{eqnarray}}
\def\eea{\end{eqnarray}}
\def\be{\begin{equation}}
\def\ee{\end{equation}}
\def\nn{\nonumber\\}
\def\tJ{t\textrm{-}J}
\def\tJV{t\textrm{-}J\textrm{-}V}
\def\hc{\mathrm{h.c.}}
\def\kag{kagom\'e }
\def\up{\uparrow}
\def\down{\downarrow}
\def\br{\mathbf{r}}
\def\bk{\mathbf{k}}
\def\bq{\mathbf{q}}
\def\bQ{\mathbf{Q}}
\def\s{\sigma}
\def\g{\gamma}
\begin{document}
\author {Martin Indergand}
\affiliation{Institut f\"ur Theoretische Physik, ETH Z\"urich, CH-8093 Z\"urich, Switzerland}

\author {Carsten Honerkamp}
\affiliation{Theoretical Physics, Universit\"at W\"urzburg, D-97074 W\"urzburg, Germany}

\author {Andreas L\"auchli}
\affiliation{Institut Romand de Recherche Num\'erique en Physique des Mat\'eriaux (IRRMA), CH-1015 Lausanne, Switzerland}

\author {Didier Poilblanc}
\affiliation{Laboratoire de Physique Th\'eorique, CNRS \& Universit\'e de Toulouse, F-31062 Toulouse, France }
  
\author {Manfred~Sigrist}
\affiliation{Institut f\"ur Theoretische Physik, ETH Z\"urich, CH-8093 Z\"urich, Switzerland}

\date{\today}
\title{Plaquette bond order wave in the quarter-filled extended Hubbard model on the checkerboard lattice}

\begin{abstract}
An extended Hubbard model (including nearest-neighbor repulsion and antiferromagnetic spin exchange)
is investigated on the frustrated checkerboard lattice, a
two-dimensional analog of the pyrochlore
lattice. Combining Gutzwiller renormalized mean-field (MF) calculations, exact diagonalization (ED) techniques, and
a weak-coupling renormalization group (RG) analysis we provide strong evidence for a crystalline valence bond plaquette 
phase at quarter-filling. The ground state is twofold degenerate and breaks translation symmetry. The bond energies 
show a staggering while the charge distribution remains uniform.
\end{abstract}
\maketitle

\section{Introduction}

Many frustrated quantum magnets exhibit complex behavior deeply rooted in the macroscopic degeneracy of their classical analog. Spin liquids characterized by the absence of any kind of symmetry breaking and valence bond crystals (VBC) characterized by translation symmetry breaking 
only (SU(2) symmetry is preserved) are exotic phases possibly realized in some two-dimensional (2D) or three-dimensional (3D) frustrated 
magnets.\cite{review} The VBC instability in 2D and 3D are higher dimensional realizations of the one-dimensional (1D) spin-Peierls mechanism 
which is known to occur, e.g., in the Heisenberg spin-1/2 chain with nearest and next-nearest  neighbor exchange interaction above some critical 
frustration (even in the absence of lattice coupling).\cite{Haldane} 
In that case, the chain spontaneously dimerizes leading to a twofold degenerate gapped ground state (GS). Similarly, the GS of the Heisenberg model on the checkerboard lattice (the square lattice with diagonal bonds on every second plaquette, see Fig.~\ref{fig:VBC}) was shown to be twofold degenerate with long range {\it plaquette} VBC order characterized by a $\bQ=(\pi,\pi)$ ordering wavevector.\cite{fouet}
In this state the spins pair up in four-spin singlet units located on every second {\it void} plaquette giving rise to stronger bonds on those plaquettes.

So far the investigation of charge degrees of freedom in geometrically frustrated
systems has been poorly explored. Most computations deal with a small fraction of holes
doped into a Mott insulator~\cite{doping} or with itinerant systems at half-filling.\cite{itinerant}

It has been suggested recently,\cite{Indergand06} that for strongly
correlated electrons on frustrated lattices such as the 2D
\kag or the 3D pyrochlore lattice a related spontaneous
symmetry breaking away from half-filling might occur at the fractional filling $n=2/3$ and $n=1/2$, respectively.
These so-called bisimplex lattices are arrays of corner sharing triangular or tetrahedral units.
Interestingly, although the GS is twofold degenerate due to the breaking of inversion symmetry that leads to two types of non-equivalent units, lattice translational symmetry is preserved
and all sites remain equivalent, i.e., there is no charge density wave (CDW) ordering. 
Each site of the lattice belongs to two non-equivalent units and
the two types of units have different bond strength, i.e., different expectation values for the kinetic and magnetic energy, which is the typical characteristic of a bond order wave (BOW).

The checkerboard lattice can be viewed as a 2D corner-sharing array of tetrahedra. As shown in Ref.~\onlinecite{Indergand06} the $\tJ$ model with two electrons on a single tetrahedron has a very robust non-degenerate and non-frustrated ground state which is given by an equal amplitude superposition of the six states having a singlet on one bond.
Therefore, it was conjectured in Ref.~\onlinecite{Indergand06} that a BOW might occur quite generally on a bipartite and corner-sharing
arrangement of tetrahedra at quarter-filling, i.e., that the BOW instability occurs not only in the 3D pyrochlore lattice but also in the 2D checkerboard lattice. 

In the present paper, we provide numerical and analytical evidence for this new type of spontaneous BOW instability for  correlated  electrons on the checkerboard lattice  at quarter-filling ($n=1/2$). The spontaneous symmetry breaking produces a doubling of the unit cell like in the spin-Peierls and VBC scenarios and does not lead to a breaking of inversion symmetry as on the pyrochlore lattice. The BOW instability occurs on the checkerboard lattice at quarter-filling
in the presence of nearest-neighbor (n.n.) interactions (repulsion and magnetic exchange).

The BOW phase bears similarities with the VBC phase at half-filling (Heisenberg model), namely long range plaquette-plaquette correlations with the same
wavevector, but the spin singlets are formed on the {\it crossed plaquettes} for the BOW at quarter-filling whereas they are formed on the {\it empty plaquettes} for VBC at half-filling.
We report the results of strong-coupling ($U=\infty$) approaches in Sec.~\ref{sec:strong}, where Sec.~\ref{sec:tJ} is dedicated to a  Gutzwiller MF calculation and Sec.~\ref{sec:ED} contains ED results.

In Sec.~\ref{sec:weak} we derive an effective weak-coupling model on the square lattice and analyze it with RG methods. The Sec.~\ref{sec:weak_coupling1} contains a MF and two-patch RG analysis and in Sec.~\ref{sec:weak_coupling2} the results of the more sophisticated $N$-patch RG are presented.

The different approaches provide a very consistent picture and show that the BOW instability on the checkerboard lattice  at quarter-filling occurs both for weak and strong coupling and for quite general repulsive interactions. 
\subsection{The Hamiltonian}
We study  the Hamiltonian
$
H=H_0+H_{\mathrm{int}}
$
at quarter-filling ($n=1/2$) on the checkerboard lattice.
The kinetic part is given by
\be
\label{Hnull}
H_0=-t\sum_{\langle ij\rangle}\sum_{\s=\up\down}\left(c^{\dagger}_{i\sigma}c^{\phantom\dagger}_{j\sigma}+\hc\right)
\ee
with the positive hopping matrix element $t$.    
The sum $\sum_{\langle ij\rangle}$  runs over all bonds on the checkerboard lattice.
The interaction part of the Hamiltonian is given by 
\be
\label{Hint}
 H_{\mathrm{int}}= U\sum_i n_{i\uparrow}n_{i\down}+ J\sum_{\langle ij\rangle}\mathbf{S}_i\cdot\mathbf{S}_j+V\sum_{\langle ij\rangle}n_i n_j
\ee
with onsite repulsion $U$, n.n.\ repulsion $V$, and n.n.\ spin exchange $J$.
For $U=\infty$ we obtain the strong coupling Hamiltonian
\bea
\label{Hstrong}
H_{\tJV}&=&\mathcal{P}H_0\mathcal{P}+J\sum_{\langle ij\rangle}
{\bf S}_i \cdot {\bf S}_j +V\sum_{\langle ij\rangle}n_i n_j,\\
\label{HstrondED}
&=&H_{\tJ}+V'\sum_{\langle ij\rangle}n_i n_j,
\eea
where $\mathcal{P}$ is the projection operator that enforces the single occupancy constraint and 
$V'=V+J/4$.
For $V'=0$ the strong coupling Hamiltonian reduces to the usual 
$\tJ$ model.

\section{Strong coupling}
\label{sec:strong}
\subsection{Gutzwiller renormalized mean-field}
\label{sec:tJ}
We study the $\tJV$ model Hamiltonian (\ref{Hstrong}) on the
checkerboard lattice. This Hamiltonian describes the physical processes in the restricted Hilbert space of configurations with no doubly occupied sites. Physically, we expect $V>J$ but
we do not necessarily assume V large compared to $t$. The local
constraints of no doubly occupied sites are replaced by
statistical Gutzwiller weights~\cite{Gut63} and a decoupling in
the particle-hole channel leads to the following MF Hamiltonian,
  \begin{eqnarray}
   H_{\rm MF}= &-& t \sum_{\langle ij\rangle}\sum_{\sigma} g_{ij}^t
      \big(c^{\dagger}_{i\sigma}c^{\phantom\dag}_{j\sigma}+\hc\big)\nonumber \nn
   &-&\sum_{\langle ij\rangle\sigma}\Big(\frac{3}{4} g_{ij}^J J + V\Big)
\big(\chi^{\phantom\dag}_{ji}c^{\dagger}_{i\sigma}c^{\phantom\dag}_{j\sigma} + \hc -|\chi_{ij}|^2\big)\nn
&+& V \sum_{\langle ij\rangle} (\langle n_i\rangle n_j
+ \langle n_j\rangle n_i -\langle n_i\rangle\langle n_j\rangle
),
  \end{eqnarray}
where the Gutzwiller weights have been expressed in terms of local
fugacities $z_i=(1-\langle n_i\rangle)/(1-\langle n_i\rangle/2)$, $g_{ij}^t=\sqrt{z_i z_j}$ and
$g_{ij}^J=(2-z_i)(2-z_j)$, to account for possible (small)
non-uniform charge modulations (if any).\cite{gs} The self-consistency
conditions are implemented as $\chi_{ji}=\langle
c^\dagger_{j\sigma}c^{\phantom\dag}_{i\sigma}\rangle$. Recently, this
approach was successfully used to investigate the properties of $4\times 4$
checkerboard-like BOW in the lightly doped $\tJ$ model
on the (non-frustrated) square lattice.\cite{4x4}

\begin{figure}
  \centerline{\includegraphics*[width=0.65\linewidth]{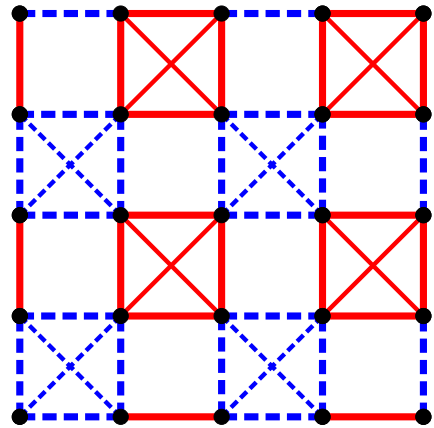}}
  \caption{\label{fig:VBC}
(Color on-line) Schematic pattern of the plaquette phase.
Different line styles (colors) correspond to different electron hopping amplitudes and spin-spin correlations (see
Fig.~\protect\ref{fig:OP}).}
\end{figure}

\begin{figure}
\includegraphics*[angle=0,width=\linewidth]{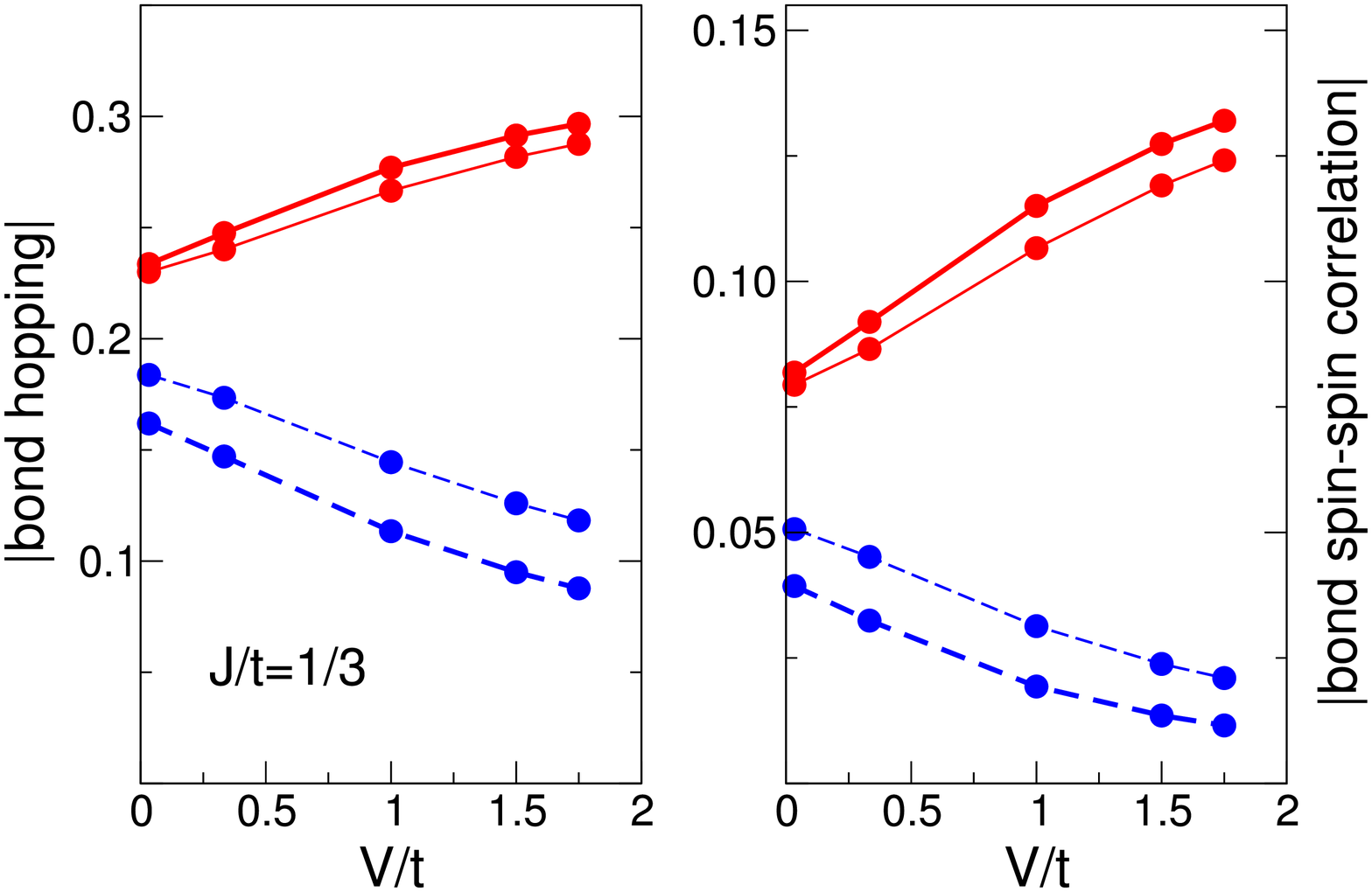}
  \caption{\label{fig:OP}
(Color on-line) Expectation values vs.\ $V/t$ of the hopping
$|\big<(c_{i\sigma}^\dagger c^{\protect\phantom\dagger}_{j\sigma} + \hc)\big>|$ (left) and exchange
$|\big< {\bf S}_i\cdot{\bf S}_j\big>|$ (right) operators on the four
non-equivalent bonds of the plaquette phase (see
Fig.~\protect\ref{fig:VBC}).
A different line style and color is used for the different plaquettes and the thicker (thiner) lines correspond to the outer (crossed) bonds of the plaquette.
The results are obtained from an {\it unrestricted} MF calculation on a $48\times 48$ lattice for $J/t=1/3$.}
\end{figure}

The MF equations are solved self-consistently on $16\times 16$ and
$48\times 48$ lattices starting from a random bond and site
configuration. We assume a fixed physical value of $J/t=1/3$ and
systematically vary the ratio $V/t$. We find that in the range
$0.3\le V/t\le 1.75$ the MF equations converge to a unique
solution given by the checkerboard pattern shown in Fig.~\ref{fig:VBC}. Interestingly, finite size effects are weak
for this range of parameters. As far as symmetry properties are
concerned, the modulated structure found here corresponds to a
doubling of the unit cell with two non-equivalent ``weak'' and
``strong'' tetrahedra (crossed plaquettes) hence leading to four
types of bonds schematically represented by different line styles (colors).
Each bond is characterized by the expectation values of the
hopping and of the spin-spin correlations which are plotted in
Fig.~\ref{fig:OP}. It is important to emphasize here that no
particular superstructure was assumed in the calculation (the
equations are solved in real space starting from a random state)
and that our solution space {\it a priori} allows for CDW's
and/or bond currents (complex values for $\chi_{ij}$). However,
our solution exhibits real values of $\chi_{ij}$ (it does not
break time-reversal symmetry) and has a uniform charge density on all sites. 
It is therefore a pure BOW. Note that two types of bonds (with slightly different amplitudes) 
are found within each tetrahedron. Although
the MF treatment, very likely, overestimates the amplitude of the
modulations,\cite{Web06} these results convincingly suggest a
quite strong BOW instability of the checkerboard lattice. Note also that
for large $V/t$ values our MF approach fails to converge as the 
system of non-linear equations becomes unstable and oscillates between two attractors.
In fact, in the strong $V/t$ limit, the ``tetrahedron
rule''~\cite{Anderson_rule} (or ice rule) should be obeyed, namely
there should be exactly two particles within each crossed
plaquette. It is clear that the simple decoupling scheme of the
$V$ term does not fulfill this ``hard-core'' constraint.

\subsection{Exact Diagonalizations}
\label{sec:ED}
\begin{figure}
 {(a)\ $N=20$}
 \vspace{2mm}$\mbox{}$
  \centerline{\includegraphics*[angle=0,width=0.6\linewidth]{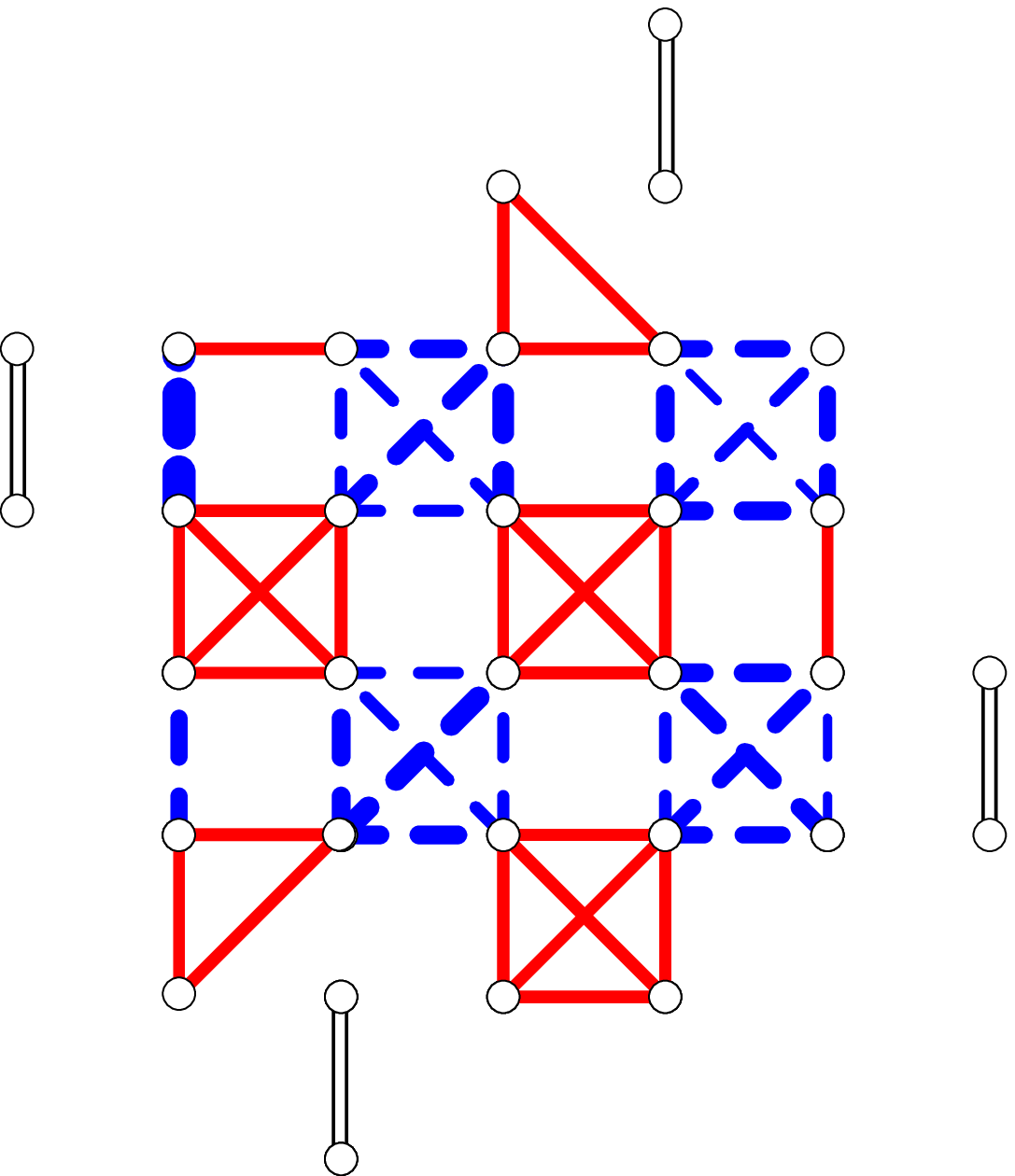}}
 $\mbox{}$\\
 \vspace{2mm}
  {(b)\ $N=24$}
 \vspace{2mm}$\mbox{}$
  \centerline{\includegraphics*[angle=0,width=0.8\linewidth]{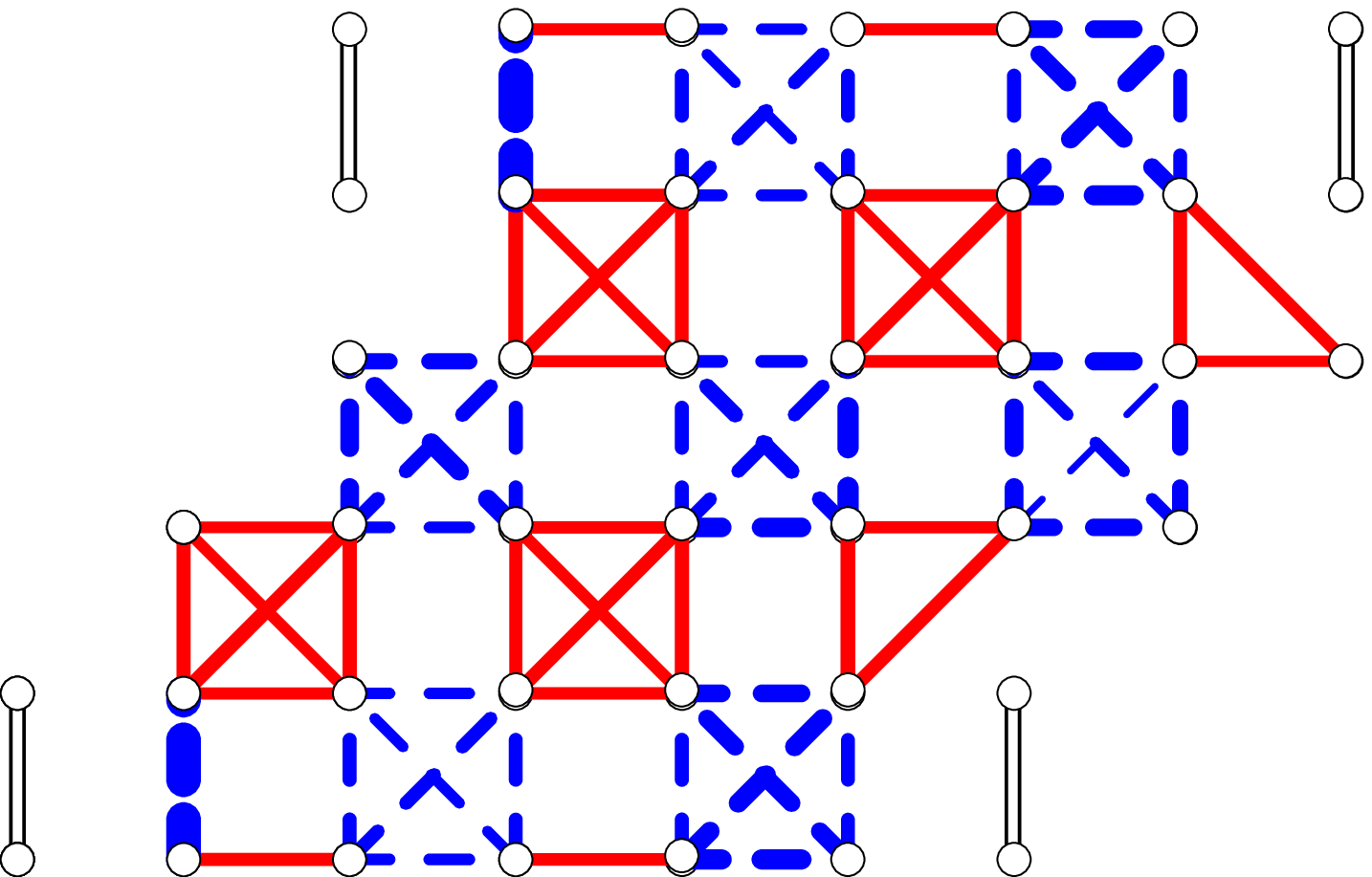}}
  \caption{
  (Color online)
  Correlation function of the kinetic energy (Eq.~\protect{\ref{eqn:kinetic_correlation}})
    of (a) a 20 sites and (b) a 24 sites checkerboard  sample at $n=1/2$, $J/t=1/3$, and $V'/t=7$.
    The black, empty bonds denote the same reference bond,
    the red, full bonds negative and the blue, dashed bonds positive
    correlations. The line strength is proportional to the magnitude of the correlations.
    \label{fig:picture_ed_bond_correlations}}
\end{figure}

We study the strong coupling $\tJV$ model (\ref{HstrondED}) within ED on
finite-size checkerboard samples of $N=16,20,$ and 24 sites at
the filling $n=1/2$.\footnote{The Hilbert space of the largest
sample has a dimension of approximately 100 million after symmetry reduction.}

In finite, periodic systems the BOW instability can be detected using a correlation
function of the bond strengths. Here we chose to work with the kinetic term. The
correlation function is defined as:
\be
  C_\mathcal{K}[(i,j),(k,l)]=\langle\mathcal{K}(i,j)\ \mathcal{K}(k,l) \rangle
 -\langle\mathcal{K}(i,j)\rangle \langle\mathcal{K}(k,l)\rangle,
  \label{eqn:kinetic_correlation}
\ee
where
\begin{equation}
\mathcal{K}(i,j)=-\ \sum_{\sigma}c^\dagger_{i\sigma}c^{\phantom\dagger}_{j\sigma}+\hc,
\end{equation}
and $(i,j)$ is a n.n.\ bond while $(k,l)$ denotes any
other bond which is linked in the Hamiltonian. We evaluate the
correlation function for all possible combinations without common
sites in the ground state of several samples at $n=1/2$ and
$J/t=1/3$, for varying $V'/t$.\footnote{%
Note, that in this section we work with the Hamiltonian \protect\ref{HstrondED} and use the parameter 
$V'=V+J/4$ instead of $V$. For $V'/t=7$ and $J/t=1/3$ the difference between $V$ and $V'$ is very 
small ($(V'-V)/V=0.012$).
}
Note that a BOW of the type
presented in Section~\ref{sec:tJ} is characterized by long range
correlations of $C_\mathcal{K}$ corresponding, in the
thermodynamic limit, to a modulation of $\langle\mathcal{K}(i,j)\rangle$ 
to be compared directly to the MF values in the left panel of Fig.~\ref{fig:OP} (the long distance 
correlation function is proportional to the magnitude-squared of the kinetic energy 
modulation).

First we plot a real space picture of the correlations on the $N=20$ and $N=24$ samples
at a value of $V'/t=7$.
As discussed below this is the value of $V'/t$ where the correlations
are strongest. The reference bond uniquely belongs to a certain class of crossed plaquettes.
Based on the theoretical picture one expects the correlation function to be positive for all
bonds on the same type of crossed plaquettes and negative on the others. This is indeed
what is seen on both samples shown in Fig.~\ref{fig:picture_ed_bond_correlations}. Note
also that the correlations are rather regular and uniform throughout the system.

\begin{figure}
  \centerline{\includegraphics*[angle=0,width=0.95\linewidth]{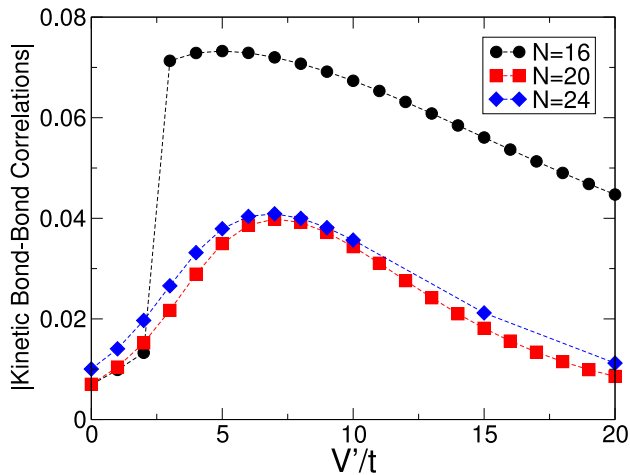}}
  \caption{
  Behavior of the absolute value of a particular bond correlation function (Eq.~\protect{\ref{eqn:kinetic_correlation}})
  at largest distance as a function of $V'/t$ at fixed $J/t=1/3$ and $n=1/2$ for several system sizes.
  Correlations are weak at small $V'/t$, peak around $V'/t\approx7$ and then decrease again towards the
  extremely large $V'/t$ limit, where another phase might appear.
  \label{fig:ed_bond_correlations}}
\end{figure}
We now track the evolution of the correlations as a function of $V'/t$ by choosing the value of the
correlation function at the largest possible distance on a given sample. We compare the
evolution on the three finite size samples in Fig.~\ref{fig:ed_bond_correlations}. The correlations
all start at small values for $V'/t\lesssim 2$, and in that region an inspection of the spatial structure
similar to Fig.~\ref{fig:picture_ed_bond_correlations} reveals that the bond correlation pattern
has defects, i.e., correlations with the wrong sign. This makes it difficult to conclude firmly on the
presence of the BOW phase at small repulsion based on the ED simulations only.
The abrupt jump seen on the $N=16$ sample is due to a level crossing, which is absent on the
larger samples.
For intermediate values of the repulsion the pattern becomes correct on all bonds and the
amplitudes peak around a repulsion of $V'_\mathrm{max}/t\approx7$ for the largest two samples.
Beyond this value the correlations weaken again, but the pattern remains qualitatively intact. The physics
at very large $V/t$, where states fulfilling the "tetrahedron rule" are dominant,  
is of particular interest but beyond the scope of this work.
Calculations for spinless fermions (i.e. polarized
electrons)~\cite{spinless} reveal a more complicated translation
symmetry breaking state with a larger supercell.
In the case of bosons however, a similar plaquette BOW is found
although involving the {\it void} plaquettes
where cyclic exchange terms act in this limit~\cite{bosons}. 
Preliminary numerical calculations and analytical considerations~\cite{spinfull}
show further that this second type of  BOW phase (not shown) is
also realized for fermions when spin degrees of freedom are included.
The investigation of the transition occuring between these two types of plaquette phases is left for future work.
\section{Weak coupling}
\label{sec:weak}
\subsection{Two-patch RG and MF analysis}
\label{sec:weak_coupling1}
We turn now to the weak-coupling analysis of the Hamiltonian
$
H=H_0+H_{\mathrm{int}}
$
with $H_0$ and $H_{\mathrm{int}}$ given by Eq.~(\ref{Hnull}) and Eq.~(\ref{Hint}), respectively.
The quadratic Hamiltonian, $H_0$, in reciprocal space 
consists of a dispersing band and a flat band lying on top (for $t>0$) of the dispersing band. If we introduce a chemical potential term with $\mu=-2t$ we can write $H_0$ in the form   
  $\sum_{\bk\sigma}\xi^{\phantom\dag}_{\bk} \g^{\dag}_{\bk\s}\g^{\phantom\dag}_{\bk\s}+4tN_{\mathrm{flat}}$ where the operator $N_{\mathrm{flat}}$ counts  the number of electrons in the flat band at $4t$. In the weak-coupling limit these states do not affect the low energy physics of the system and therefore they will be dropped in the following analysis.
 The dispersion of the other band is given by $\xi_{\bk}=-2t(\cos{k_1}+\cos{k_2})$ and is identical to the tight-binding dispersion on the square lattice. It is particle-hole symmetric, perfectly nested with the nesting vector $\bQ=(\pi,\pi)$, and has a logarithmically divergent density of states at the Fermi energy. 

\begin{figure}
  \centerline{\includegraphics*[angle=0,width=0.95\linewidth]{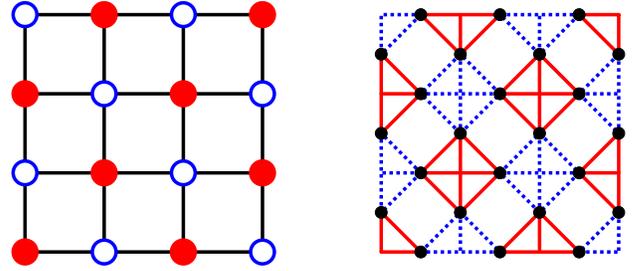}}
  \caption{\label{fig:sCDW}
(Color on-line) Correspondence between the $s$-CDW phase on the effective square lattice (left) and the plaquette phase or BOW phase on the original checkerboard lattice (right).}
\end{figure}

For small couplings $U$, $V$, and $J$ we obtain an effective interaction Hamiltonian  (Appendix~\ref{app:weakcoupling})
given by
\begin{equation}
\label{Heffective}
  H_{\mathrm{Int}}^{\mathrm{eff}}=\frac{1}{2N}\sideset{}{'}\sum_{\bk_1\dots\bk_4}g^{\phantom\dag}_{\bk_1\dots\bk_4}\,
  \sum_{\sigma\sigma'}\g^{\dag}_{\bk_1\sigma}\g^{\dag}_{\bk_2\sigma'}\g^{\phantom\dag}_{\bk_3\sigma'}\g^{\phantom\dag}_{\bk_4\sigma},
\end{equation}
with
\bea
\label{eq:gkkkk}
g^{\phantom\dag}_{\bk_1\dots\bk_4}&=&\sum_{\nu=1}^2
(Ue^\nu_{\bq}+2\tilde{V}e^{\nu}_{\bq_{23}}-Je^{\nu}_{\bq_{24}})
e^{\nu}_{\bk_1}e^\nu_{\bk_2}e^\nu_{\bk_3}e^\nu_{\bk_4}\nn
&+&\sum_{\nu=1}^2 (4\tilde{V}f^{\nu}_{\bk_1\bk_4}f^{\bar{\nu}}_{\bk_2\bk_{3}}-2Jf^{\nu}_{\bk_1\bk_3}f^{\bar{\nu}}_{\bk_2\bk_4})
\eea
with $e^{\nu}_{\bk}=\cos(k_\nu/2)$, $f^{\nu}_{\bk\bk'}=e^{\nu}_{\bk-\bk'}e^{\nu}_{\bk} e^{\nu}_{\bk'}$,  $\tilde{V}=V-J/4$, $\bq=\bk_1+\bk_2-\bk_3-\bk_4$, and $\bq_{ij}=\bq-2(\bk_i-\bk_j)$.
The two saddle points $P_1=(0,\pi)$ and $P_2=(\pi,0)$ of the dispersion lead to the logarithmic divergence of the density of states.
Therefore, we can characterize very weak interactions by the values of the function $g_{\bk_1\dots\bk_4}$ where all four momenta $\bk_1\cdots\bk_4$ lie on one of the two saddle points.
Adopting the notation of Ref.~\onlinecite{binz} we have the following four coupling constants:
\be
g_{\bk_1\dots\bk_4}=\left\{ \begin{array}{ll}
g_1=-2J                  &\bk_1,\bk_3\in P_1\\
g_2=+4\tilde{V}       &\bk_1,\bk_4\in P_1\\
g_3=0                      &\bk_1,\bk_2\in P_1\\
g_4=U+2\tilde{V}-J  &\bk_1\dots\bk_4\in P_1
\end{array}\right.
\ee
From the RG equations of Ref.~\onlinecite{binz} we see that for positive values of $U$, $V$, and $J$  the coupling $g_1$ flows to $-\infty$ and the coupling $g_2$ flows to $+\infty$,
\footnote{For $\tilde{V}<0$ $g_2$ flows to 0, but also in this case the CDW phase is stabilized.}
 whereas the other couplings flow to 0.
This shows that the charge density wave susceptibility is diverging most rapidly under the RG flow. 
As shown in Fig.~\ref{fig:sCDW} the CDW instability on the square lattice corresponds to the BOW instability on the checkerboard lattice, i.e., both instabilities break exactly the same symmetries.
Note, that within the framework of this RG scheme, we can not determine whether a usual CDW or a so-called charge flux phase, which is a CDW with a $d$-wave form factor, is stabilized. In the following we will denote these two phases with $s$-CDW and $d$-CDW.
 (Note, that for $J=0$ only the coupling $g_2$ diverges. In this case we can not even determine whether a CDW of SDW phase is stabilized.)
In order to show, that the $s$-CDW phase is in fact favored over the $d$-CDW, at least in a mean-field analysis, we restrict the interaction Hamiltonian (\ref{Heffective}) to the CDW channel and obtain:
\be
H_{\mathrm{CDW}}=\frac1{4N}\sum_{\bk\bk'}V_{\bk\bk'}\sum_{\sigma\sigma'}\g^{\dag}_{\bk\sigma}\g_{\bk+\bQ,\sigma}\g^{\dag}_{\bk'+\bQ,\sigma'}\g_{\bk'\sigma'}
\ee
with
\bea
V_{\bk\bk'}
&=&2g_{\bk,\bk'+\bQ,\bk',\bk+\bQ}-g_{\bk,\bk'+\bQ,\bk+\bQ,\bk'}\\
&=&V^s_{\bk\bk'}+V^d_{\bk\bk'}+V^{d'}_{\bk\bk'}+V^p_{\bk\bk'}
\eea
{\small
\bea
\label{VCDWs}
V^s_{\bk\bk'}&=&-V_0\left[(1-\cos k_1\cos k_2)(1-\cos k_1'\cos k_2')   \right.\\
&& +\left. (\sin^2k_1+\sin^2 k_2)(\sin^2k_1'+\sin^2k_2')/2\right]\nn
\label{VCDWd}
V^d_{\bk\bk'}&=&-V_0(\cos k_1-\cos k_2)(\cos k'_1-\cos k'_2)\\
\label{VCDWdp}
V^{d'}_{\bk\bk'}&=&-V_0\sin k_1\sin k_2\sin k_1'\sin k_2' \\
\label{VCDWp}
V^p_{\bk\bk'}&=&+\frac{U}{4}(\sin k_1 \sin k_1'+\sin k_2 \sin k_2')
\eea}%
where $V_0=(\tilde{V}+J)/2$. In (\ref{VCDWd}) we dropped a term proportional to $\cos k_1+\cos k_2$, as it vanishes along the FS, and in  (\ref{VCDWp}) we dropped terms proportional to $V/U$ or $J/U$.
With $\Delta_{\bk}=\frac1N\sum_{\bk'}V_{\bk\bk'}F_{\bk'}$ and $F_{\bk}=\sum_\sigma \langle \gamma^{\dag}_{\bk+\bQ,\sigma}\gamma^{\phantom\dag}_{\bk\sigma}\rangle$ we obtain the linearized gap equation
\be
\Delta_{\bk}=-\frac{1}{N}\sum_{\bk'}V_{\bk,\bk'} \frac{\tanh(\xi_{\bk'}/2T_\mathrm{c})}{2\xi_{\bk'}}\Delta_{\bk'}.
\ee
Note, that the $d'$- \eqref{VCDWdp} and the $p$-wave \eqref{VCDWp} instability do not open a gap at the saddle points, furthermore, the $p$-wave instability is strongly repulsive.
For the $d$-CDW state the pair potential (\ref{VCDWd}) is separable and the linearized gap equation can be written in the simpler form
\be
\label{Tcd}
1=\frac{V_{0}}{N}\sum_{\bk}(\cos k_1-\cos k_2)^2 \frac{\tanh(\xi_{\bk}/2T^d_{\mathrm{c}})}{2\xi_{\bk}}.
\ee
%
The pair potential for the $s$-CDW state (\ref{VCDWs}) is not separable. But if we neglect for a moment the second line in (\ref{VCDWs}) we obtain an analogous expression to (\ref{Tcd}) for the critical temperature of the $s$-CDW with $\cos k_1-\cos k_2$ replaced by $1-\cos k_1\cos k_2$.  As on the FS we have $\cos k_2=-\cos k_1$ and as $|2\cos k_1|\leq 1+\cos^2 k_1$ we know that the critical temperature of the $s$-CDW phase is higher than the critical temperature of the $d$-CDW phase.  Including the non-negative second term in (\ref{VCDWs}) would only lead to a further increase of the critical temperature.
Note, that the $d$- and the $s$-CDW potentials are identical for the momenta on the saddle points. Therefore it is not surprising that they cannot be distinguished by the two-patch RG method. However, in the mean-field we can see that the $s$-CDW is favored over the $d$-CDW state, as it opens a finite gap along the entire FS.
It is possible to perform such a mean-field analysis also for SDW and superconducting instabilities. For superconductivity we have strong repulsion ($\propto U$) in the $d$- and in the $s$-wave channel, and  weaker ($\propto V,J$) and mainly repulsive interactions in the $d'$- an the $p$-wave channel. It is therefore quite clear, that the superconducting instabilities can not compete with the $s$-CDW instability.
For the pair potential of the SDW instabilities, we should replace only $V_0$ by $\tilde{V}/2$ and $U$ by $-U$ in Eqs.
~(\ref{VCDWs}-\ref{VCDWp}). For $J>0$ we have $V_0=(\tilde{V}+J)/2>\tilde{V}/2$ and therefore the $s$-, $d$- and $d'$-CDW instabilities are favored over the corresponding SDW instabilities.
The  $p$-SDW instability is strongly attractive but has the handicap, that it does not open a gap at the saddle points, therefore, for weak interactions the $s$-CDW state will still be favored over the $p$-SDW state, i.e., for interaction parameters $(U,V,J)$ given by $(\alpha u,\alpha v,\alpha j)$ we can find for all positive values of $(u,v,j)$ an $\alpha_0$ such that for $0<\alpha<\alpha_0$ the $s$-CDW state is stabilized.

In conclusion, in this section we showed with RG and mean-field arguments, that for weak enough repulsive and antiferromagnetic interactions the $s$-CDW phase, which corresponds to a bond order wave  state on the checkerboard lattice, is stabilized.

\subsection{$N$-patch functional RG method}
\label{sec:weak_coupling2}
As the two-patch model in the last section could not determine whether the
$s$- or the $d$-CDW ground state is favored, we advance to a more refined
$N$-patch functional renormalization group (fRG) study of the effective model
for the dispersive band. This allows us to study the whole Fermi surface and,
in particular, the competition between $s$-CDW and $d$-CDW state.
The effective interactions in these channels and also the
bare effective interaction possess a non-trivial $\bk$-dependence outside the saddle point regions which can be taken into account in the $N$-patch fRG scheme.

We set up a one-loop $N$-patch fRG calculation of the one-particle irreducible interaction vertex using the methods described in Refs.~\onlinecite{hsfr}. This fRG scheme integrates out all intermediate one-loop processes (particle-hole diagrams and particle-particle contributions) with a decreasing energy scale $\Lambda$. In the one-loop diagrams at scale $\Lambda$, one intermediate particle is in the energy shell with band energy $\pm \Lambda$, while the other particle is further away from the Fermi surface. The RG flow is started at $\Lambda_0 \sim$ bandwidth. The integration of the one-loop processes in the flow down to lower $\Lambda$ renormalizes the wavevector-dependence of the effective interaction. From the growth of certain components toward low scales, the dominant correlations can be read off. 
In parallel with the running interactions, the flow of various static
susceptibilities can be obtained. These quantities are again renormalized by
the scale-dependent interactions via one-loop processes. Their growth can be
compared as a function of the RG scale. In the case of a divergence of the
interactions at a nonzero critical scale $\Lambda_{\mathrm{c}}$, the fasted growing
susceptibility determines the dominant instability or ordering tendency 
of the system.  
 
For the spin-rotationally invariant system, the scale-dependent interaction vertex can be expressed using a running coupling function $V_\Lambda(\bk,\bk', \bk+\bq)$. 
The initial coupling function, $V_{\Lambda_0}(\bk,\bk', \bk+\bq)$, is obtained from  Eq.~\eqref{eq:gkkkk} as $g_{\bk+\bq,\bk'-\bq,\bk',\bk}$. 
Here the frequency dependence has been neglected as in most previous applications of the method for correlated fermions.\cite{frgrevs} In order to describe the wavevector dependence, we use 32 angular patches around the Fermi surface.\cite{zanchi} These patches are again split up in 3 patches, one above the Fermi surface down to band energies of $0.2t$, one including the Fermi surface in the band energy window between $-0.2t$ and $0.2t$, and one below this energy window. We have checked our results using other partitions. 
\begin{figure}
\includegraphics[width=.485\textwidth]{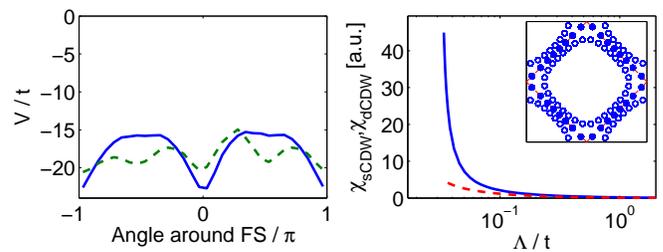}
\caption{fRG results for a $32\times 3$ discretization of the Brillouin zone,
  for the half-filled band, $T=0.001t$, $\tilde{V}=0.6t$, $J=0.4t$ and $U=0$. 
Left: Effective interactions $V_\Lambda (\bk,\bk',\bk+\bQ)$ with $\bQ=
(\pi,\pi)$ and $\bk$ near the $(\pi,0)$-point at angle $0$ (solid line)
or near the $(\pi/2,\pi/2)$-point at angle $\pi/4$ (dashed line) vs.\ $\bk'$
moving around the FS at scale $\Lambda=0.33t$ where the strongest attractive component of the coupling function exceeds $-24t$.
Right: Flow of the zero-frequency $s$-CDW (solid line) and $d$-CDW susceptibility for the same parameters. The inset shows the 32 FS points (filled circles) and the additional patch centers at band energy $\pm 0.4t$ (open circles).
}
\label{frgfig1}
\end{figure}

We have run the fRG for half band filling and various choices of the
interaction parameters $U$, $\tilde V$ and $J$ in (\ref{Heffective}). From a perturbative point of
view, the model with pure onsite repulsion and $\tilde V=J=0$ is remarkably
stable. 
For instance at $U=2t$ and half filling of the dispersive band, the fRG flow of the
interactions remains finite down to low scales and temperatures $\sim 10^{-4}\ t$
without any indication for an instability. This stability is quite surprising
for a perfectly nested Fermi surface with van Hove singularities at the Fermi
level. However, as already found in the two-patch model, for many scattering processes within the dispersive band
between states near the van Hove points, the bare interactions vanish, and the
interactions away from these points and away from the Fermi level (which are
kept in approximate form in our refined patching scheme) 
are apparently not sufficient to produce an instability at a higher scale.

For nonzero nearest-neighbor interactions $\tilde V$ and $J$ the fRG produces
run-away flows at measurable scales. 
For instance, for $U=0$, $\tilde V = 0.6t$ and $J =0.4t $ we find a flow to strong coupling, i.e., a divergence of some components of $V_\Lambda(\bk,\bk', \bk+\bq)$ in the one-loop flow, at scales $\Lambda_{\mathrm{c}} \approx 0.033t$. 
This scale sets an upper temperature bound for possible long range order. The divergence is dominated by scattering processes $V_\Lambda (\bk,\bk',\bk+\bQ)$ with $\bk$ and $\bk'$ at the FS and with wavevector transfer $\bQ=(\pi,\pi)$ which flow to strongly negative values. 
For incoming and outgoing particles near the saddle points, these processes belong to the $g_1$-type scattering known to diverge to $- \infty$ from the previous section. In the two-patch model, also the $g_2$-type processes seemed to diverge. This tendency is not seen in the fRG for the full Fermi surface, now we observe only a small growth of the $g_2$-type processes.    
In Fig. \ref{frgfig1} we display the diverging component $V_\Lambda (\bk,\bk',\bk+\bQ)$ of the effective interactions for two different choices of $\bk$ vs.\ $\bk'$ moving around the FS at a low scale $\Lambda$ where the maximal component of the coupling function exceeds $-24t$. 
 One clearly observes that the data for $\bk$ fixed near $(\pi,0)$ is basically a
 superposition of an attractive offset and a $\cos k_1 - \cos k_2$
 modulation. The attractive offset becomes more negative with decreasing scale
 and drives the $s$-CDW susceptibility. The $d$-wave part also grows and feeds
 the $d$-CDW channel. For $\bk$ fixed near the BZ diagonal, i.e., near the nodes
 of the $\cos k_1 - \cos k_2 $ form factor, the $d$-wave part is missing or
 strongly deformed, while the attractive offset is more or less
 unchanged. Hence we can expect a stronger growth in the $s$-CDW channel. This
 is also clearly reflected in the data for the flow of the susceptibilities
 shown in the right panel of Fig.~\ref{frgfig1}. The growth of the $s$-CDW
 susceptibility by far exceeds the $d$-CDW channel and all other channels such
 as spin density waves. Hence the fRG treatment agrees with the meanfield
 picture developed in the previous section. Both suggest a $s$-CDW ordered
 ground state in the effective one-band model, translating into a bond-order
 wave as primary order in the full checkerboard system.  

The $s$-CDW divergence described here occurs in a rather large parameter window. We increased the value of $U$ from 0 up to $2t$ and found an increase of the critical scale from $0.033t$ to $0.037t$ together with an even stronger dominance of the $s$-CDW over the $d$-CDW channel. 

\section{Conclusions}

We studied the Hubbard model including n.n.\ repulsion and n.n.\ antiferromagnetic exchange for strongly correlated electrons at quarter-filling ($n=1/2$) on the checkerboard lattice in the strong-coupling ($U=\infty$) and in the weak-coupling regime.
The MF calculations, the ED simulations and the RG analysis all predict a spontaneous breaking of the translation symmetry 
in the ground state, that is characterized by the nesting vector $\bQ=(\pi,\pi)$. The ground state is twofold degenerate as there are two types
of crossed plaquettes (``strong'' and ``weak'' plaquettes).
These ground states can be approximatively described as product states of localized two-particle states on the ``strong'' plaquettes, which are equal amplitude superpositions of a singlet wave function on each of the six bonds.
In this way the two electrons forming a singlet gain the magnetic  energy $-J$ and the kinetic energy $-4t$.  Furthermore, the presence of a third electron on the crossed plaquette which would cost a n.n.\ repulsion energy $2V$ is avoided.
Therefore, the BOW instability is a cooperative effect that results from a simultaneous optimization of kinetic and interaction energy.
Note, that for negligible kinetic energy ($V\gg t$) other phases can be realized on the quarter-filled checkerboard lattice.%
\footnote{%
As an example consider the case $t=0$ and $U=V=\infty$, where the constraint of two particles per crossed plaquette allows only states where occupied sites form closed (or infinite) loops on the checkerboard lattice.
For $J=0$ all these states are degenerate but for finite $J$ the states, where all loops have the minimal length four, are ground states, as the magnetic energy per electron for the loops of length four  is minimal.
 The closed loops of four electrons around open plaquettes can be placed in a periodic pattern where the centers of the loops are connected by the vectors $2\mathbf{x}_1,2\mathbf{x}_2$ (cf.\ App.~\protect\ref{app:weakcoupling}).
Note, that in this periodic arrangement all four electron plaquettes on a diagonal can simultaneously slide along a diagonal line by $\mathbf{x}_1$ or $\mathbf{x}_2$ at no energy cost. Therefore, the ground state is in fact infinitely degenerate. 
}
An interesting issue for further theoretical studies are the properties and the excitations of the weakly doped BOW phase, e.g., the question whether the additional holes or particles pair and lead to a superconducting phase where the U(1) gauge symmetry and the translation symmetry are simultaneously broken.

In conclusion with the discovery and the description of the BOW phase on the quarter-filled checkerboard lattice we show how a strongly frustrated system can avoid the local frustration at fractional filling by exploiting the bipartite arrangement of larger units.
We hope that with our work we can stimulate further progress in the challenging and exciting field of strongly correlated electrons on frustrated lattices.  

\acknowledgments

We thank T.M.~Rice, K.~Wakabayashi, K.~Penc, and S.~Capponi for
stimulating discussions. We acknowledge support by the Swiss
National Fund and NCCR MaNEP (Switzerland).
Computations were performed on the IBM Regatta machines of CSCS
Manno and IDRIS (Orsay).
D.P. thanks the Agence Nationale de la Recherche (France) for support and the ITP at ETH-Z\"urich for hospitality.

\appendix
\section{checkerboard lattice\label{checkappRG}}
\label{app:weakcoupling}

The elementary unit cell of the checkerboard lattice contains two lattice sites situated at $\mathbf{x}_\nu/2$
 ($\nu=1,2$) where the vectors $\mathbf{x}_\nu$ are two primitive lattice vectors.
 With this convention we choose the origin of the lattice at the center of a
 crossed plaquette,
 where there is no lattice site.
 The tight-binding Hamiltonian for this system is given by
 \begin{eqnarray}
   \label{checktb}
 H_0&=&-t\sum_{\br\nu\s}\left[c^{\dag}_{\br\nu\s}(c^{\phantom{\dag}}_{\br+\mathbf{x}_{\nu}
 ,\nu\s}+c^{\phantom{\dag}}_{\br-\mathbf{x}_{\bar{\nu}},\bar{\nu}\s}\right.\\
 &&\left.\qquad\qquad\mbox{}+c^{\phantom{\dag}}_{\br-\mathbf{x}_{1}+\mathbf{x}_{\nu},\bar{\nu}\s}
 )+\hc\right]-\mu \hat{N},\nonumber
 \end{eqnarray}
 where $\bar{\nu}=2,1$ if $\nu=1,2$, $\hat{N}$ is the number operator and $\mu=-2t$ in the following.
 We introduce Fourier transformed operators as
 \begin{equation}
 \label{ftcheck}
 c_{\br\nu\s}=\frac{1}{\sqrt{N}}\sum_{\bk}e^{i\bk\cdot(\br+\frac{\mathbf{x}_{
 \nu}}{2})}c_{\bk\nu\s},
 \end{equation}
 where $N$ is the number of unit cells.
 With these operators the tight-binding Hamiltonian reads
 \begin{equation}
   \label{checktbft}
 \hat{H}_0=4t\sum_{\bk\mu\nu\s}\Big[\delta_{\mu\nu}-\cos\big(\frac{k_{\mu}}{2}
 \big)\cos\big(\frac{k_{\nu}}{2}\big)\Big]c^{\dag}_{\bk\mu\s}c^{\phantom{\dag}}_{\bk\nu\s},
 \end{equation}
 where $k_\nu=\bk\cdot\mathbf{x}_\nu$.
 Diagonalizing this Hamiltonian leads to a flat band at $4t$ and to a band with the dispersion
  $\xi_{\bk}=-2t\sum_\nu\cos k_\nu$
 which is nothing but the nearest-neighbor tight-binding dispersion of the square lattice.
 The operators of this dispersive band are expressed in terms of the original operators by
 \begin{equation}
   \label{ccheck}
 \g_{\bk\s}=\frac{1}{\sqrt{r_{\bk}}}\sum_{\nu}\cos\big(\frac{k_{\nu}}{2}\big)\,c_{
 \bk\nu\s}
 \end{equation}
 with $r_{\bk}=\sum_{\nu}\cos^2(k_{\nu}/2)$.
In weak coupling, we can restrict our attention to the states close to the Fermi surface, where $r_{\bk}=1$ and for every operator on the checkerboard lattice we can obtain an effective operator on the square lattice by the substitution  $c_{\bk\nu\s}\rightarrow \cos(k_{\nu}/2)\g_{\bk\sigma}$.
In this way the effective Hamiltonian (\ref{Heffective}) is obtained from the interaction Hamiltonian (\ref{Hint}).

\end{document}